\documentclass[aps,prd,showpacs,showkeys,nofootinbib,preprintnumbers,twocolumn,longbibliography,superscriptaddress]{revtex4-1}
\pdfoutput=1

\usepackage[final]{epsfig}
\usepackage{amsmath}
\usepackage{amssymb}
\usepackage[hidelinks]{hyperref}
\usepackage{times}
\usepackage{dsfont}
\usepackage{xfrac}
\usepackage{bm}
\usepackage{slashed}
\usepackage[usenames,dvipsnames]{xcolor}
\usepackage[utf8]{inputenc}
\usepackage[english]{babel}

\input cyracc.def
\font\tencyr=wncyr10
\def\cyr{\tencyr\cyracc}

\renewcommand{\vec}[1]{\boldsymbol{\mathbf{#1}}}

\newcommand{\V}{{\rm V}}
\newcommand{\A}{{\rm A}}
\newcommand{\GF}{G_{\rm F}}
\newcommand{\pF}{p_{\rm F}}

\newcommand{\half}{{\textstyle\frac{1}{2}}}
\def\bfrac#1#2{\displaystyle\frac{#1}{#2}}

\definecolor{blue}{rgb}{0,0,0.5}
\definecolor{georg}{rgb}{0,0,0.9}
\definecolor{lightblue}{rgb}{0,0,1}
\definecolor{alex}{rgb}{0.5,0,0}
\definecolor{lightred}{rgb}{1,0.5,0}
\definecolor{green}{rgb}{0,0.5,0}
\definecolor{darkgreen}{rgb}{0.0,0.3,0.0}
\definecolor{orange}{rgb}{1,0.4,0}
\definecolor{grey}{rgb}{0.5,0.5,0.5}

\begin{document}

\preprint{MPP-2016-88}

\title{Helicity oscillations of Dirac and Majorana neutrinos}

\author{Alexandra Dobrynina}
\affiliation{P.~G.~Demidov Yaroslavl State University, Sovietskaya 14, 150003 Yaroslavl, Russia}
\affiliation{Max-Planck-Institut f\"ur Physik (Werner-Heisenberg-Institut),
F\"ohringer Ring 6, 80805 M\"unchen, Germany}

\author{Alexander Kartavtsev}
\author{Georg Raffelt}
\affiliation{Max-Planck-Institut f\"ur Physik (Werner-Heisenberg-Institut),
F\"ohringer Ring 6, 80805 M\"unchen, Germany}

\begin{abstract}
The helicity of a Dirac neutrino with mass $m$ evolves under the
influence of a $B$-field because it has a magnetic dipole moment
proportional to $m$.  Moreover, it was recently shown that a polarized
or anisotropic medium engenders the same effect for both Dirac and
Majorana neutrinos. Because a \hbox{$B$-field} polarizes a background
medium, it instigates helicity oscillations even for Majorana
neutrinos unless the medium is symmetric between matter and
antimatter. Motivated by these observations, we review the impact of a
$B$-field and of an anisotropic or polarized medium on helicity
oscillations for Dirac and Majorana neutrinos from the common
perspective of in-medium dispersion.
\end{abstract}

\maketitle

\section{\label{sec:Introduction}Introduction}

The left-handed nature of weak interactions implies that astrophysical
neutrino ensembles emerge far from helicity equilibrium, notably in
the early universe and the interior of collapsing stars or merging
neutron stars. On the other hand, the ``wrong-helicity'' components
are not completely sterile because neutrinos have small
masses. Moreover, there could be new interactions which could
accelerate the relaxation toward left-right equilibrium, although we
will here focus on the impact of neutrino masses alone. In the latter
case, the rate towards helicity equilibrium is of order $(m/2E)^2$
times an ordinary weak rate, where $E$ is a typical neutrino
energy. This rate tends to be far too small to be of practical
interest. However, the helicity conversion rate can be coherently
enhanced in the form of helicity oscillations.

It has long been known that nonzero neutrino masses imply magnetic and
electric dipole and transition moments \cite{Fujikawa:1980yx,
  Pal:1981rm, Shrock:1982sc, Nieves:1981zt, Schechter:1981hw,
  Dobrynina:2014rza, Giunti:2014ixa}, allowing $B$-fields to
instigate spin and spin-flavor oscillations
\cite{Werntz:1970, Cisneros:1970nq, Lim:1987tk, Akhmedov:1987nc}. Only
Dirac neutrinos have intrinsic magnetic dipole moments, whereas both
Dirac and Majorana neutrinos have magnetic and electric transition
moments originating from the mismatch between mass and interaction
eigenstates.

It was only recently fully appreciated that analogous effects arise
for both Dirac and Majorana neutrinos if the background medium is
anisotropic either in the form of a convective current (background
vector current) or polarized (background axial vector current)
\cite{Vlasenko:2013fja, Cirigliano:2014aoa, Vlasenko:2014bva,
  Volpe:2013jgr, Vaananen:2013qja, Serreau:2014cfa,
  Kartavtsev:2015eva}. The origin of this effect is the mismatch
between chirality and helicity for neutrinos with small masses in
analogy to the origin of the traditional dipole moment. A medium
polarization normally originates from a $B$-field which therefore can
flip the spin both by the intrinsic dipole moment and indirectly by
polarizing the medium. The interaction energy of both effects must be
of order $e B\GF m$, so it is not obvious which is more important.

One way to look at this phenomenon is the perspective that in a medium
both a Dirac and a Majorana neutrino acquires an effective magnetic
moment: the spin polarization of the medium caused by the $B$-field is
interpreted as an in-medium electromagnetic vertex function. A
significant body of literature has studied this point to determine the
dispersion law of active neutrinos and antineutrinos, but usually not
with an eye for helicity evolution \cite{Semikoz:1987py,
  Semikoz:1989za, Nieves:1989ez, Nieves:1989xg, D'Olivo:1989cr,
  Semikoz:1994uy, Elmfors:1996gy, Nunokawa:1997dp, Erdas:1998uu,
  Egorov:1999ah, Grigoriev:2002zr, Studenikin:2004bu,
  Kuznetsov:2005tq}. We here take the opposite approach and interpret
even the normal $B$-field as yet another helicity-changing refractive
medium and not in terms of intrinsic dipole moments.

The purpose of our short review is to look at neutrino helicity
flipping by a background medium ($B$-field, unpolarized medium with
currents, polarized medium) from a common perspective and to provide
explicit expressions for different cases, considering both Dirac and
Majorana neutrinos with equal or different masses and considering a
medium with a large matter-antimatter asymmetry as well as one which
is matter-antimatter symmetric. Most of these results can be found
scattered in the literature, but we hope to provide a useful
clarification by unifying them from a common perspective.

We begin in Sec.~\ref{sec:self-energy} with general aspects of
neutrino dispersion and discuss how anisotropic media lead to helicity
conversion when neutrinos have Dirac or Majorana masses. In
Sec.~\ref{sec:Bfield} we turn specifically to a $B$-field and show
that the general formulas coincide with the usual description in terms
of dipole moments. In Sec.~\ref{sec:media} we consider an electron gas
polarized by a $B$-field and compare with the $B$-field-only case.  We
conclude in Sec.~\ref{sec:conclusions}.

\section{\label{sec:self-energy}Helicity conversion}

In this section we study generic aspects of neutrino dispersion and
helicity evolution in homogeneous and static media or $B$-fields on
the level of refraction (forward scattering).
Of course, in this setup
a nontrivial evolution arises only if the initial state does not
coincide with one of the in-medium propagation eigenstates.

\subsection{Neutrino dispersion}

The medium causes a shift of the neutrino self-energy $\Sigma$ by a
background-induced term $\Sigma_{\rm b}$. In the Dirac equation in
Fourier space, it appears in the form
\begin{equation}
\left(\slashed{k}-m-\Sigma_{\rm b}\right)\psi=0\,,
\end{equation}
where $k$ is the neutrino four-momentum,
$\slashed{k}=k_\mu\gamma^\mu$, and $m$ the neutrino mass. Generally
$\Sigma_{\rm b}$ has a nontrivial Dirac structure and depends on $k$
besides medium properties. Nontrivial solutions require ${\rm
  det}\left(\slashed{k}-m-\Sigma_{\rm b}\right)=0$, equivalent to
asking for the poles of the propagator. This condition determines the
dispersion relation of the propagation eigenstates
  in the given background medium.

\begin{figure}[!t]
\includegraphics[width=0.80\columnwidth]{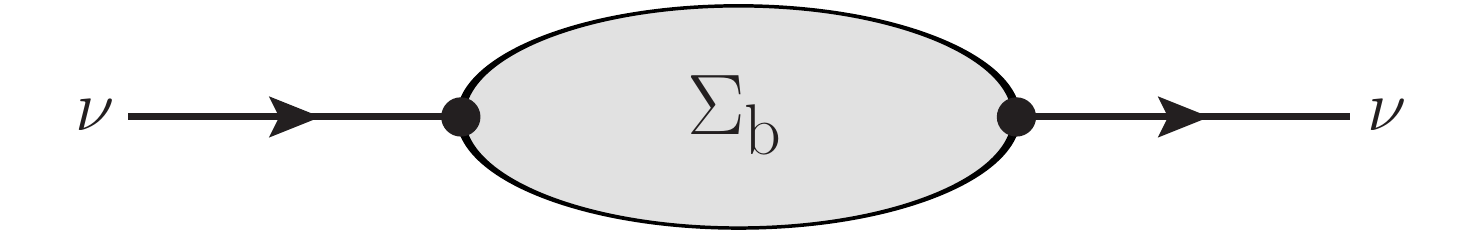}
\vskip2pt
\caption{Neutrino self-energy graph in a background medium or
  electromagnetic field.}
\label{fig:blob0}
\vskip20pt
\includegraphics[width=0.80\columnwidth]{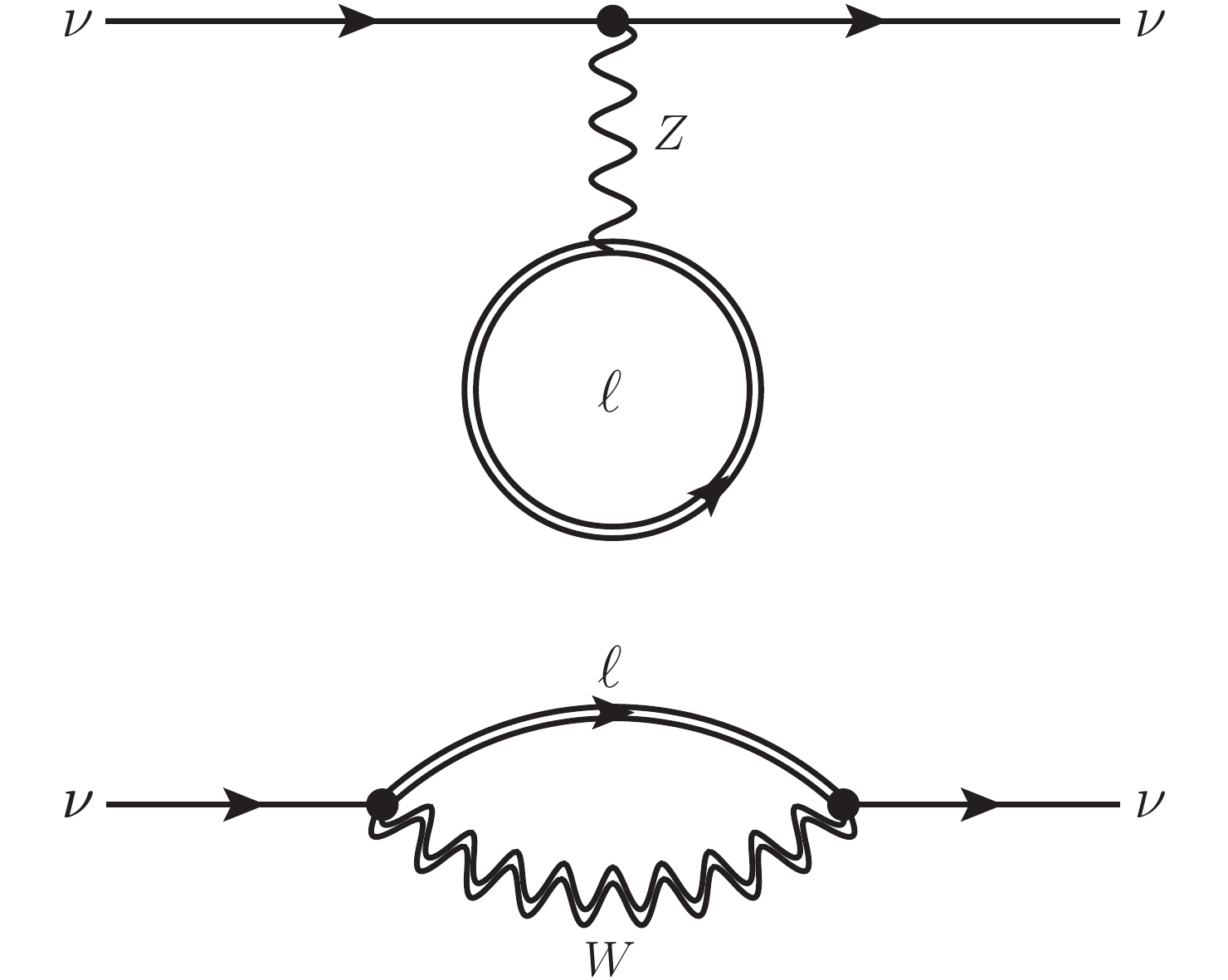}
\vskip2pt
\caption{One-loop neutrino self-energy (unitary gauge) in a medium or
  $B$-field. {\em Top:}~Tadpole graph. {\em Bottom:}~Charged-current
  bubble graph. Double lines are dressed propagators, i.e., they
  include real particles of the medium and real or virtual states in
  the $B$-field.  In our physical situations of interest, $W$ and $Z$
  bosons are always virtual.  We ignore background neutrinos,
  otherwise a $Z$ bubble would also appear. 
  The absence of background neutrinos renders the
  evolution equations linear and we can study each momentum
  mode separately.
\label{fig:blob1}}
\end{figure}

The neutrino self-energy is the ``blob'' (the truncated matrix
element) in Fig.~\ref{fig:blob0}. In unitary gauge to one-loop order
and ignoring background neutrinos, $\Sigma_{\rm b}$ is given by the
tadpole and bubble graphs of Fig.~\ref{fig:blob1} which must involve
the chirality projections\footnote{In the Russian literature,
  $\gamma_5$ is defined with opposite sign and thus appears in the
  definitions of $R$ and $L$ with opposite sign.}
$R=\frac{1}{2}(1+\gamma_5)$ and $L=\frac{1}{2}(1-\gamma_5)$, where
neutrino lines are attached.  The only Dirac structure surviving the
chirality projections is $\gamma^\mu$ or $\gamma^\mu\gamma_5$, the
latter contributing a negative sign that is absorbed in the common
coefficient. Therefore, on one-loop level we arrive at the most
general form
\begin{equation}\label{eq:self-energy-form}
\Sigma_{\rm b}=-R\left(a \slashed{k}+\slashed{b}\right)L\,,
\end{equation}
where $a$ is a dimensionless coefficient. The overall sign follows the
convention of Ref.~\cite{Weldon:1982aq}. The four-vector $b$ of
dimension energy depends on the background currents
  and fields. It may also involve $k$, except for a term proportional
to $k$ which we have explicitly separated.

For massless neutrinos, the dispersion of  active and
sterile components is independent. The sterile components, if they exist,
are unaffected by the medium. For the active components, the coefficient
$a$ modifies $\slashed{k}\to(1+a)\slashed{k}$ in the Dirac equation
and one finds\footnote{We use the metric $(+,-,-,-)$ so that $J^0$ and
  $J_0$ of a four-vector $J$ are the same. For typographical
  simplicity we usually write $J_0$ for the time-like component. In an
  expression like $J=(J_0,{\bm J})$ the three-vector ${\bm J}$ refers
  to the contravariant space-like components, i.e., to $J^i$ with
  $i=1$, 2 or 3. The Lorentz invariant product of two four-vectors is
  $IJ=I_\mu J^\mu=I_0J_0-{\bm I}\cdot{\bm J}$.}
\begin{equation}
\label{eq:spectrum}
k_0+\frac{b_0}{1+a}=\pm\left|{\bm k}+\frac{{\bm b}}{1+a}\right|\,.
\end{equation}
Henceforth we will neglect $a\ll 1$ because perturbatively it will contribute
only higher-order corrections.

This result is not as simple as it looks because $b$ can depend on
$k$.  However, in the ultra-relativistic limit, where dispersion
effects are small, we can use the unperturbed $k$ to express $b$. Even
then we must worry about the sign of $k_0$, where negative-energy
solutions of the Dirac equation represent positive-energy
antiparticles with opposite momentum. Henceforth we will write the
dispersion relation for positive energies for both neutrinos and
antineutrinos. After expanding to linear order in $b$ we arrive at
\begin{equation}
\label{eq:diracdispersion}
k_0=|{\bm k}|-(b_0-b_{\parallel})\times
\begin{cases}
1&\hbox{for $\nu$,}\\
\eta_b&\hbox{for $\bar\nu$,}
\end{cases}
\end{equation}
where \smash{$b_\parallel\equiv \hat{\bm k} \cdot {\bm b}$} and
$\hat{\bm k}$ is a unit vector in the direction of $\bm k$. The
component $b_\perp$ transverse to the neutrino momentum does
not affect neutrino dispersion in linear approximation.

The parameter $\eta_b=\pm1$ in
  Eq.~\eqref{eq:diracdispersion} had to be introduced as a price for
  interpreting antineutrinos as positive-energy states. In particular,
$\eta_b=-1$ when $b$ does not depend on $k$ and therefore the
dispersion effect has opposite sign for $\bar\nu$. This is the case
for an ordinary medium with large matter-antimatter asymmetry. On the
other hand, $\eta_b=+1$ when the dispersion effect does not change
sign. Typical examples are a matter-antimatter symmetric medium or a
$B$-field in vacuum.

Generally there are contributions to $b$ from
different types of background, e.g., a medium of different components
and a $B$-field.  Because Eq.~(\ref{eq:diracdispersion}) for the
ultra-relativistic limit is linear in $b$ we can consider each
component separately.

\subsection{Helicity evolution}

Neutrinos have small masses so that the ultrarelativistic dispersion
relation in vacuum is $k_0=|{\bm k}|+m^2/2|{\bm k}|$.  Together with
flavor mixing, it leads to the usual flavor
conversion effects. In addition, if ${\bm b}$ has a nonvanishing
component transverse to ${\bm k}$, transitions between
states of opposite helicity occur to linear order in the masses.
Moreover, the background effect is diagonal in the weak-interaction
basis which is very different from the mass basis. Therefore,
spin-flavor transitions are also possible.

\begin{figure}[!t]
\includegraphics[width=0.8\columnwidth]{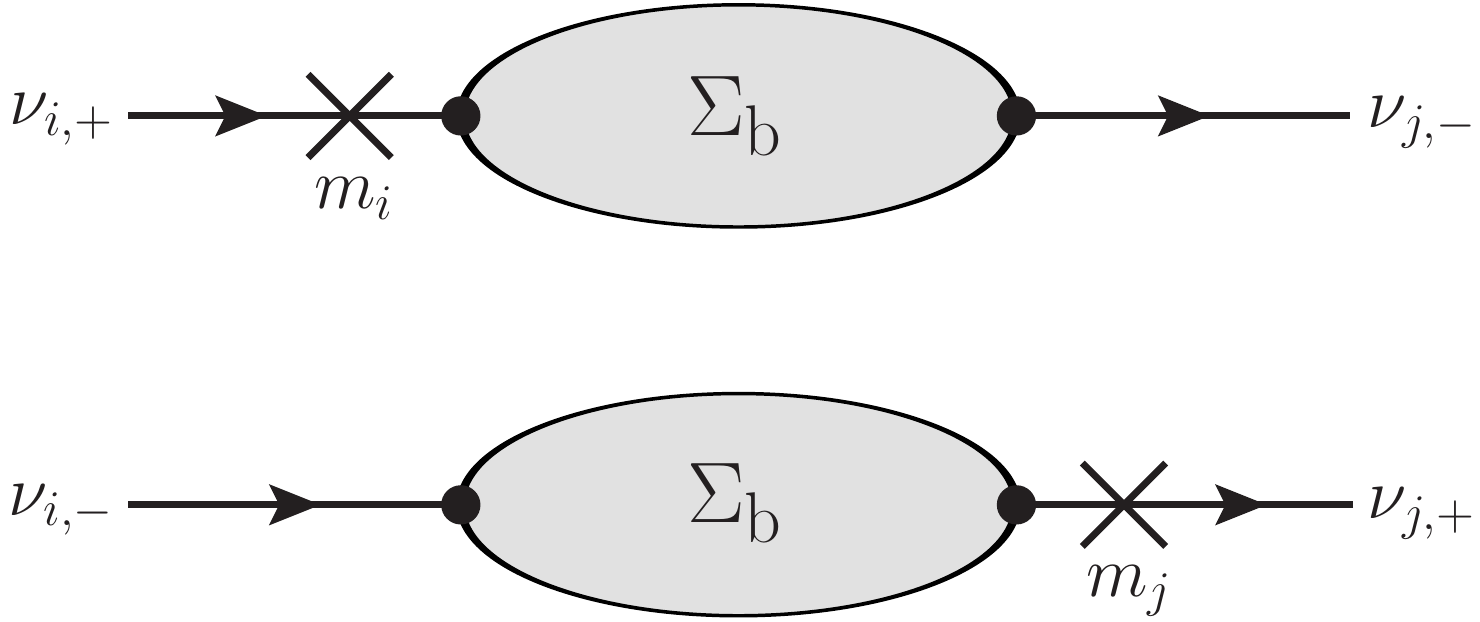}
\vskip2pt
\caption{Spin-flip amplitudes between Dirac neutrinos with masses
  $m_i$ and $m_j$. The amplitude is proportional to the mass of the
  ``wrong-helicity'' state attached to the blob which represents the
  in-medium self-energy of left-handed neutrinos.} \label{fig:blob2}
\end{figure}

To understand helicity conversion qualitatively, we use the mass basis
and denote the eigenstates with $i$ or $j$. Without calculation we
glean from Fig.~\ref{fig:blob2} how the masses enter
when we are in the ultrarelativistic
limit. For Dirac neutrinos, which possess four
  degrees of freedom just like the charged leptons, only the
l.h.\ neutrino components communicate with the self-energy blob, so it
is the mass of the ``wrong-helicity'' (sterile) Dirac state which
intervenes to connect to $\Sigma_{\rm b}$. In particular, the
amplitude for \smash{$\nu_{i,+}\to\nu_{j,-}$} is proportional to
$m_i$, while for \smash{$\nu_{i,-}\to\nu_{j,+}$} it is proportional to
$m_j$. If the two masses are identical, the amplitude is proportional
to $m=m_i=m_j$ and therefore the same for both cases.

To calculate the helicity-flipping amplitude it is easiest to begin
with the exact spinors describing massive neutrinos in vacuum
\cite{Serreau:2014cfa}.  A Dirac neutrino with mass $m_i$, momentum
${\bm k}$, and helicity $h=\pm1$ is described by a spinor
$u_{i,h}({\bm k})$. The forward-scattering effect of the medium is
encoded in the energy shifts represented by
  $\Sigma_{\rm b}$.  In the basis of free massive neutrinos, as opposed to
  the in-medium propagation eigenstates considered earlier, the shift
of the neutrino Hamiltonian~is
\begin{equation}
\label{eq:Hnunu}
{\sf H}^{\nu\nu}_{ij,sh}=\bar u_{i,s}\Sigma_{\rm b}u_{j,h}\,,
\end{equation}
where the spinors are taken at momentum ${\bm k}$ and we use the
same normalization \smash{$u_{i,h}^\dagger u_{i,h}=1$}
  as adopted in Refs.~\cite{Serreau:2014cfa, Kartavtsev:2015eva}. A
similar construction applies to antineutrinos with 
${\sf H}^{\bar\nu\bar\nu}_{ij,sh}$ using $v$-spinors
\cite{Kartavtsev:2015eva}.  The self-energy has the simple form
Eq.~(\ref{eq:self-energy-form}).  As explained earlier, to lowest
order we can restrict ourselves to the $b$ term, so that
\begin{equation}
\label{eq:Hnunu-2}
{\sf H}^{\nu\nu}_{ij,sh}=-(\bar u_{i,s}R)\slashed{b}(Lu_{j,h})\,.
\end{equation}
In agreement with our qualitative arguments presented earlier,
$L u_{j,+}\propto m_j$ and $\bar{u}_{i,+}R\propto m_i$, so that
$H_{ij,-+}\propto m_j$ and $H_{ij,+-}\propto m_i$.

Dispersion based on interaction with a medium is diagonal in the
weak-interaction basis and we denote with $b_\ell$ the contribution
related to the charged lepton $\ell$, see Fig.~\ref{fig:blob1}. For
the transition between mass states $i$ and $j$, the relevant
background charged current is therefore
\begin{equation}
b_{ij}^\mu=\sum_{\ell=e,\mu,\tau} U^\dagger_{i \ell}\,b_\ell^\mu\, U_{\ell j}\,,
\end{equation}
where $U$ is the unitary leptonic mixing matrix. Neutral-cur\-rent
interactions are diagonal in both the flavor and mass basis because of unitarity of
the neutrino mixing matrix,
\begin{equation}\label{eq:unitarity}
\sum_{\ell=e,\mu,\tau} U^\dagger_{i\ell}\,U_{\ell j}=\delta_{ij}\,,
\end{equation}
and therefore do not contribute to flavor or spin-flavor transitions.
If neutrino-neutrino interactions are neglected then the different
momentum modes decouple.  Without loss of generality we may then
choose a coordinate system such that for each mode the neutrino momentum
is in the $z$-direction, whereas the background current lies in
the $x$--$z$--plane. We arrange the elements of the matrix ${\sf H}^{\nu\nu}_{ij,sh}$
in such a way that the equation of motion can be cast in the form
\cite{Kartavtsev:2015eva}
\begin{equation}
\label{eq:spinflavorstruct}
i\partial_t
\begin{pmatrix}\,\framebox{\hbox to 0pt{$\vphantom{+}$}$-$}_{~i}\\[1ex]
\,\framebox{\hbox to 0pt{$\vphantom{+}$}$+$}_{~i}\end{pmatrix}=
\begin{pmatrix}
\framebox{\hbox to 0pt{$\vphantom{+}$}$--$}_{~ij}&\framebox{$-+$}_{~ij}\\[6pt]
\framebox{$+-$}_{~ij}&\framebox{$++$}_{~ij}
\end{pmatrix}
\begin{pmatrix}\,\framebox{\hbox to 0pt{$\vphantom{+}$}$-$}_{~j}\\[1ex]
\,\framebox{\hbox to 0pt{$\vphantom{+}$}$+$}_{~j}\end{pmatrix},
\end{equation}
where each box denoted by helicities $\pm$ 
is either a column or a $3\times 3$ matrix in flavor space. For Dirac
neutrinos we then find to linear order in neutrino masses \cite{Kartavtsev:2015eva}
\begin{equation}
\label{eq:Dirac-flavor}
  {\sf H}^{\nu\nu}_{ij,sh}=-
  \begin{pmatrix}b^0_{ij}-b^\parallel_{ij}&~b_{ij}^\perp\,\bfrac{m_j}{2|\vec{k}|}\\[2ex]
  b_{ij}^\perp\,\bfrac{m_i}{2|\vec{k}|}&0\end{pmatrix}.
\end{equation}
Indeed, the masses on the off-diagonals are the ones associated with
the ``wrong'' (positive) helicity Dirac neutrino.  For Dirac
antineutrinos, a similar expression pertains for which we find
\cite{Kartavtsev:2015eva}
\begin{equation}
\label{eq:anti-Dirac-flavor}
  {\sf H}^{\bar\nu\bar\nu}_{ij,sh}=\eta_b
  \begin{pmatrix}0&  b_{ij}^\perp\,\bfrac{m_i}{2|\vec{k}|}\\[2ex]
  b_{ij}^\perp\,\bfrac{m_j}{2|\vec{k}|}&-(b^0_{ij}-b^\parallel_{ij})\end{pmatrix}.
\end{equation}
Now the negative helicity is the sterile one and the mass\-es
appear accordingly. Notice also that the nonzero diagonal
elements of Eqs.~\eqref{eq:Dirac-flavor} and~\eqref{eq:anti-Dirac-flavor}
match the energy shifts in Eq.~\eqref{eq:diracdispersion}, as anticipated
from invariance of the eigenvalues with respect to the choice of basis.

\begin{figure}[!t]
\includegraphics[width=0.8\columnwidth]{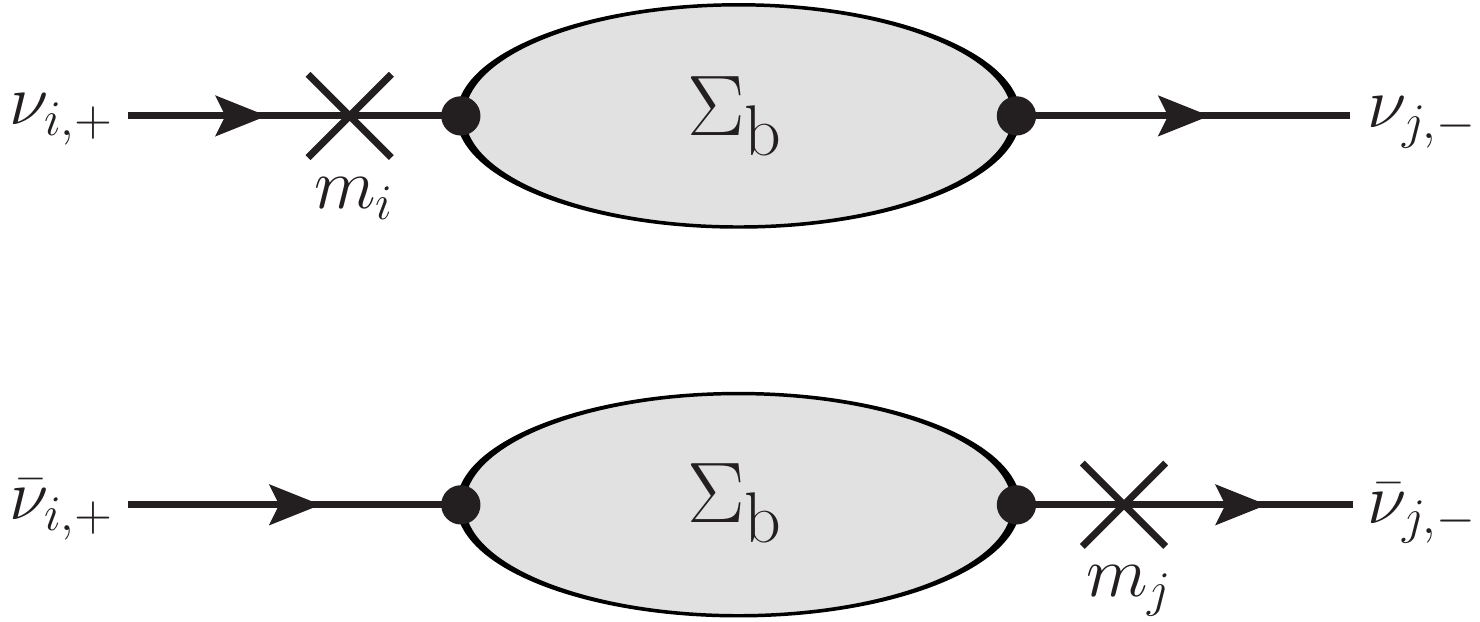}
\vskip2pt
\caption{\label{fig:blob3}Spin-flip amplitudes between Majorana
  neutrinos with masses $m_i$ and $m_j$. The positive-helicity state
  can be interpreted as a ``wrong-helicity'' neutrino or as a
  ``correct-helicity'' antineutrino. Both amplitudes
  contribute with an important relative phase.}
\end{figure}

\eject

Majorana neutrinos have only two degrees of
freedom and no helicity is wrong---both of them are active
states, essentially $\nu$ and $\bar\nu$.  Therefore, the
amplitude for $\nu_{i,+}\to\nu_{j,-}$ involves both graphs in
Fig.~\ref{fig:blob3}. For equal masses $m=m_i=m_j$, the amplitude is
proportional to $m-\eta_b m$. If the refractive effect changes sign
for $\bar\nu$ as in an ordinary medium ($\eta_b=-1$), a suitable
anisotropy engenders helicity conversions. In the opposite case where
$\eta_b=+1$ there is no helicity evolution. Typical examples are a
CP-symmetric plasma or a $B$-field in vacuum, corresponding to the
absence of a magnetic dipole moment for a Majorana neutrino.
For unequal masses $m_i$ and $m_j$, the amplitude is proportional to
$m_i-\eta_be^{i\alpha}m_j$, where $\alpha$ is a relative phase which
depends on the neutrino mixing matrix.  Typically both Dirac and
Majorana neutrinos show spin-flavor oscillations in any type of
anisotropic medium, and in particular, both have transition dipole
moments engendered by masses and flavor mixing.

To render these qualitative arguments more quantitative, we recall
that for Majorana particles \smash{$\bar{\nu}=\nu^T C$}, where $C$ is
the operator of charge conjugation. If the mean-field Hamiltonian
on the level of the current-current approximation  contains a
term $\sum_{ij}\bar{\nu}_i R\slashed{b}_{ij} L\nu_j$, it can be identically
rewritten as
\begin{equation}
\sum_{ij}\bar{\nu}_i R\slashed{b}_{ij} L\nu_j  = -\sum_{ij}\bar{\nu}_i L\slashed{b}_{ji} R\,\nu_j\,.
\end{equation}
In the language of Dirac neutrinos, the $R\slashed{b}L$ structure
projects out the neutrino contributions, whereas $L\slashed{b}R$
projects out the antineutrino contributions. Therefore, for
Majorana neutrinos the resulting mean-field Hamiltonian can be
expressed as a difference of the Dirac neutrino and antineutrino
Hamiltonians in the form
\begin{equation}\label{eq:Majorana-flavor}
  {\sf H}^{\rm M}_{ij,sh}=-
  \begin{pmatrix}
  b^0_{ij}-b^\parallel_{ij}&\bfrac{b^\perp_{ij}m_j-\eta_b b^\perp_{ji} m_i}{2|\vec{k}|}\\[2ex]
  \bfrac{m_i b^\perp_{ij}-\eta_b  m_j b^\perp_{ji}}{2|\vec{k}|}
  &\eta_b\bigl(b^0_{ji}-b^\parallel_{ji}\bigr)
  \end{pmatrix}.
\end{equation}
As expected, the diagonal elements again match the
energy shifts of the active neutrino and antineutrino respectively.

We emphasize again that these results
  apply to any kind of background medium. The Dirac vs.\ Majorana
  question is the same in all cases.

\subsection{\label{sec:OrdMedium}Unpolarized medium}

To illustrate these points let us consider the case of an ordinary
medium consisting of electrons, protons and neutrons, that has been
studied in detail in Refs.~\cite{Serreau:2014cfa,Kartavtsev:2015eva}.
Neutrino dispersion is given, at low energies, by the usual
currrent-current interaction. For a given background species, it leads
to the dispersion effect
\begin{equation}\label{eq:matter-current}
b^\mu=-\sqrt{2}\GF\,\bigl(C_\V J_\V^\mu-C_\A J_\A^\mu\bigr)\,,
\end{equation}
where $C_\V$ and $C_\A$ are vector and axial-vector couplings which depend on the
background species and the flavor of the test neutrino.  The
mean-field currents of the background fermions (field $\psi$) are
\smash{$J_\V^\mu=\langle\overline\psi\gamma^\mu\psi\rangle$} and
\smash{$J_\A^\mu=\langle\overline\psi\gamma^\mu\gamma_5\psi\rangle$}.

In the remainder of this subsection we consider an unpolirazed medium,
$J_\A=0$. If it is isotropic in its rest frame (${\bm J}_V=0$)
dispersion is given in terms of the usual weak potential
\begin{equation}\label{eq:weakpotential}
  V=\sum_{e,p,n} C_\V J_\V^0=\sqrt{2}\GF\times
\begin{cases}
(N_e-N_{\bar e})-\half N_n&\hbox{for $\nu_e$,}\\
-\half N_n& \hbox{for $\nu_{\mu,\tau}$,}
\end{cases}
\end{equation}
where $N_{e}$, $N_{\bar e}$ and $N_n$ are the electron, positron and
neutron number densities. As usual, the neutral-current proton and
electron vector contributions cancel in a neutral medium. For
antineutrinos as test particles, the potential changes sign so that
$\eta_b=-1$.
	
In the laboratory frame, the medium may flow with a velocity
${\bm\beta}$ so that $b=-V\,(1,{\bm \beta})$. Considering a single
neutrino generation with mass $m$ we find for Dirac neutrinos,
antineutrinos, and Majorana neutrinos,
\begin{subequations}
\begin{eqnarray}
i\partial_t \begin{pmatrix}\nu_-\\[1ex] \nu_+\end{pmatrix}
&=&V \begin{pmatrix}1-\beta_\parallel& \beta_\perp \bfrac{m}{2|\vec{k}|}\\[1ex]
\beta_\perp \bfrac{m}{2|\vec{k}|}&0\end{pmatrix}
\begin{pmatrix}\nu_-\\[1ex] \nu_+\end{pmatrix},
\\[2ex]
i\partial_t \begin{pmatrix}\bar\nu_-\\[1ex] \bar\nu_+\end{pmatrix}
&=&V\begin{pmatrix}0& \beta_\perp \bfrac{m}{2|\vec{k}|}\\[1ex]
\beta_\perp \bfrac{m}{2|\vec{k}|}&-(1-\beta_\parallel)\end{pmatrix}
\begin{pmatrix}\bar\nu_-\\[1ex] \bar\nu_+\end{pmatrix},
\\[2ex]
i\partial_t \begin{pmatrix}\nu\\[1ex] \bar\nu\end{pmatrix}
&=&V\begin{pmatrix}1-\beta_\parallel& \beta_\perp \bfrac{m}{|\vec{k}|}\\[1ex]
\beta_\perp \bfrac{m}{|\vec{k}|}&-(1-\beta_\parallel)\end{pmatrix}
\begin{pmatrix}\nu\\[1ex] \bar\nu\end{pmatrix}\,.
\end{eqnarray}
\end{subequations}
Note that the transverse current of an ordinary medium engenders a
nontrivial helicity evolution even for Majorana neutrinos.
Spin-flavor oscillations are given by similar equations with
the more complicated flavor structure of
Eqs.~(\ref{eq:Dirac-flavor})--(\ref{eq:Majorana-flavor}).

Whereas at low energies neutrinos interact with the mean field of an
ordinary medium in the usual form of a current-current Hamiltonian,
the medium of the early universe is nearly matter-antimatter symmetric
and the current-current refractive effect vanishes. Refractive effects
are still engendered by the nonlocal structure of the interaction,
i.e., at higher order in inverse gauge boson masses. If the medium is
isotropic, the contribution of electrons and positrons to $\nu_e$
dispersion, see Eqs.~\eqref{eq:diracdispersion},
\eqref{eq:Dirac-flavor} and \eqref{eq:anti-Dirac-flavor}, is
\cite{Notzold:1987ik}
\begin{equation}\label{eq:symmetric-bubble}
b_0=\frac{8\sqrt{2}\GF k_0}{3m_W^2}
\bigl(\langle E_e\rangle N_e+\langle E_{\bar e}\rangle N_{\bar e}\bigr)\,.
\end{equation}
Here, $E_e$ is the energy of background electrons, and $E_{\bar e}$ of
positrons. In this example, $b_0$ is proportional to the neutrino
energy $k_0$ so that the dispersion effect is the same for
$\nu_e$ and $\bar\nu_e$, in our convention meaning that $\eta_b=+1$.

\section{\label{sec:Bfield}Magnetic field}

As another specific case we now turn to neutrinos propagating in a
$B$-field in vacuum. Usually this situation is
described in terms of neutrino dipole moments induced by their masses,
but we should be able to obtain the same results if we think of the
$B$-field as a background medium affecting neutrinos in analogous ways
as a general anisotropic medium.

\subsection{Background current}

The relevant self-energy graph at
one-loop order in unitary gauge is shown in Fig.~\ref{fig:blob1}
(bottom). The double lines stand for the dressed propagator of the
virtual particles in the external $B$-field. To linear order
in $B$ and for a given charged lepton $\ell$ in the loop and
neutrino four-momentum $k$ one finds \cite{Pal:1981rm, Shrock:1982sc, Dobrynina:2014rza}
\begin{equation}\label{eq:B-field-current}
b^\mu_\ell= \frac{6e\sqrt{2}\GF}{(4\pi)^2}\,f(r_\ell)\,k_\alpha\widetilde F^{\alpha\mu}\,,
\end{equation}
where $\widetilde F^{\mu\nu}=\frac{1}{2}\epsilon^{\mu\nu\alpha\beta}F_{\alpha\beta}$
is the dual electromagnetic field-strength tensor
which in the laboratory frame is assumed to be a pure $B$-field. In
this frame, the Lorentz structure is explicitly
\begin{equation}\label{eq:bfieldstructure}
k_\alpha\widetilde F^{\alpha\mu}=-({\bm B}\cdot{\bm k},k_0{\bm B})\,.
\end{equation}
The charged-lepton dependent factor depends on the mass ratio
$r_\ell=(m_\ell/m_W)^2$ in terms of the function
\begin{equation}\label{eq:mass-exapansion}
f(r)=\frac{2-5r+r^2}{2\,(1-r)^2}
-\frac{r^2\ln r}{(1-r)^3}=1-\frac{r}{2}+{\cal O}(r^2)\,.
\end{equation}
For all charged leptons, $r_\ell\ll1$ so that we may always use the
lowest nontrivial expansion in $r_\ell$. Therefore, overall we
find
\begin{equation}\label{eq:Bfield-vacuum}
b_\ell= -\frac{6e\sqrt{2}\GF}{(4\pi)^2}\,\left(1-\frac{r_\ell}{2}\right)\,
({\bm k}\cdot{\bm B},k_0{\bm B})\,,
\end{equation}
which indeed has the dimension of energy.

\subsection{Helicity flip}

The components of $b^\mu$ include $b_0\propto |{\bm k}| B_\parallel$,
$b_\parallel\propto k_0 B_\parallel$, and $b_\perp\propto k_0
B_\perp$. In the Hamiltonian matrices
Eqs.~(\ref{eq:Dirac-flavor})--(\ref{eq:Majorana-flavor}), the
refractive effect on the diagonal is always proportional to
$b_0-b_\parallel$ which is now proportional to
$(|{\bm k}|-k_0)\,B_\parallel$ and thus of order $(m^2/2E)\,B_\parallel$.
Therefore, to linear order in neutrino masses, $B$-fields in vacuum do
not produce a refractive effect for neutrinos moving
parallel to ${\bm B}$.

To obtain the helicity-changing elements for equal masses $m=m_i=m_j$
we observe that $b$ is proportional to $k$, implying that the
refractive effect is the same for $\nu$ and $\bar\nu$ and hence
$\eta_b=+1$. Therefore,
Majorana neutrinos do not suffer helicity evolution. For Dirac
neutrinos, on the other hand,
\begin{equation}
\label{eq:Dirac-dipole-moment}
{\sf H}^{\nu\nu}_{ii,+-}=\frac{3e\sqrt{2}\GF}{(4\pi)^2}\,m_i B_\perp
\end{equation}
where the multiplier of $B_\perp$ is recognized as the usual magnetic
dipole moment for massive Dirac neutrinos. Notice that $E$ in the
denominator of $m/2E$ has canceled against $k^0$ in $b_\perp$, so
indeed the helicity-flip element does not depend on the neutrino
energy.

For transitions between different mass eigenstates, the term which is
independent of the charged fermion drops out because of the unitarity
of the neutrino mixing matrix shown in
Eq.~(\ref{eq:unitarity}). Therefore, the first nonvanishing
contribution comes from the second term in the expansion
Eq.~(\ref{eq:mass-exapansion}). With
\begin{equation}\label{eq:Fij-define}
F_{ij}=-\frac{1}{2}
\sum_{\ell=e,\mu\tau} U^\dagger_{i\ell}\biggl(\frac{m_\ell}{m_W}\biggr)^2 U_{\ell j}\,,
\end{equation}
we thus find for Dirac neutrinos
\begin{equation}
{\sf H}^{\nu\nu}_{ij,+-}=\frac{3e\sqrt{2}\GF}{(4\pi)^2}\,m_i\,F_{ij}B_\perp\,.
\end{equation}
For the opposite helicity flip, instigated by ${\sf
  H}^{\nu\nu}_{ij,-+}$, we get the other mass $m_j$, i.e., of the
participating sterile Dirac component. For Majorana neutrinos, the
corresponding element is
\begin{equation}\label{eq:Majorana-effective-dipole}
{\sf H}^{\rm M}_{ij,+-}
=\frac{3e\sqrt{2}\GF}{(4\pi)^2}
\left(m_i F_{ij}-m_j F_{ji}\right)B_\perp\,.
\end{equation}
As anticipated earlier, the amplitudes of
$\nu_i\leftrightarrow\bar\nu_j$ and $\bar\nu_i\leftrightarrow\nu_j$
are proportional to the two participating masses added with a relative
phase which depends on the details of the complex Majorana mixing
matrix. Note also that the flavor-diagonal diagonal elements of
Eq.~\eqref{eq:Majorana-effective-dipole} automatically vanish.

\subsection{Dipole moments}

We have formulated the impact of a $B$-field in vacuum on neutrino
propagation in the same way as general neutrino dispersion in a
medium. Usually, however, the same physics is described in terms of
intrinsic neutrino dipole moments which lead to spin precession in
external fields. To make contact with this more common language, we
notice that the Lorentz structure \smash{$k_\alpha\widetilde F^{\mu\alpha}$} for
the $B$-field background current given in Eq.~(\ref{eq:B-field-current})
leads to the structure \smash{$k_\alpha \widetilde F^{\mu\alpha}\gamma_\mu L$}
for the self-energy. Using common identities for Dirac matrices we can rewrite
the latter as
\begin{align}
\label{kFgamma}
k_\alpha \widetilde F^{\mu\alpha}\gamma_\mu L  = & -\textstyle{\frac{1}{8}}
(\slashed{k}\sigma_{\mu\nu}+\sigma_{\mu\nu}\slashed{k})F^{\mu\nu}\nonumber\\
& +\textstyle{\frac{1}{8}}
(\slashed{k}\gamma^5\sigma_{\mu\nu}-\gamma^5\sigma_{\mu\nu}\slashed{k}) F^{\mu\nu}\,.
\end{align}
Sandwiching the rhs.\ of Eq.~\eqref{kFgamma} between $\bar{u}_{i,s}$
and $u_{j,h}$ and using the unperturbed Dirac equation on one of the
external neutrinos, one obtains the equivalent structure
\begin{align}
-\textstyle{\frac{1}{8}} [(m_i+m_j)\sigma_{\mu\nu} F^{\mu\nu}
+(m_j-m_i)\gamma^5\sigma_{\mu\nu}] F^{\mu\nu}\,.
\end{align}
\eject
\noindent
Comparing this with the traditional dipole Lagrangian,
\begin{equation}
\label{eq:mu-Lagrangian}
{\cal L}= -\half\bar\psi_i\left(\mu_{ij}\sigma_{\mu\nu}
+i\epsilon_{ij}\gamma_5\sigma_{\mu\nu}\right)\psi_j\,F^{\mu\nu}\,,
\end{equation}
we can immediately read off the usual flavor structure of the magnetic
and electric dipole moments of Dirac neutrinos~\cite{Shrock:1982sc},
\begin{equation}
\label{eq:transitionmoments}
\begin{matrix}\mu_{ij}\\[1ex] i\epsilon_{ij}\end{matrix}\biggr\rbrace
=\frac{m_j\pm m_i}{2}\,\frac{3e\sqrt{2}\GF}{(4\pi)^2}
\left(\delta_{ij}+F_{ij}\right)\,,
\end{equation}
where the lower sign refers to $\epsilon_{ij}$.
$F_{ij}$ was defined in Eq.~(\ref{eq:Fij-define}) in terms of
the lepton mixing matrix.

Turning this argument around, in the ultrarelativistic limit
the $\gamma_5$ term in Eq.~(\ref{eq:mu-Lagrangian}) simply contributes
a sign, depending on the helicity of the initial neutrino.
Therefore, the spin precession engendered by $B_\perp$ corresponds
to  an effective transition moment of the form $\mu_{{\rm eff},ij}=\mu_{ij}\pm i\epsilon_{ij}$.
With Eq.~(\ref{eq:transitionmoments}), this amounts to
\begin{equation}
\mu_{{\rm eff},ij}=\left(\frac{m_j+m_i}{2}\pm\frac{m_j-m_i}{2}\right)
\,\frac{3e\sqrt{2}\GF}{(4\pi)^2}\,F_{ij}\,.
\end{equation}
The signs of the masses in the magnetic and electric dipole
moments work out such that the amplitude is proportional to
the mass of the ``wrong-helicity'' neutrino participating in the
spin-flavor process in agreement with our earlier discussion.

In the diagonal case ($i=j$), the coefficients $\mu=\mu_{ii}$ and
$\epsilon=\epsilon_{ii}$ are the usual magnetic and electric dipole
moments. Note that the latter automatically vanishes in the one-loop
approximation. In the neutrino rest frame, the nonrelativistic
reduction corresponds to a Hamiltonian describing a two-level system
in the form $-(\mu{\bm B}+\epsilon{\bm E})\cdot{\bm\sigma}$, where
${\bm\sigma}$ is a vector of Pauli matrices. If the neutrino spin is
not aligned with the respective electromagnetic field, it precesses
around the field direction with frequency $2\mu B$ or $2\epsilon E$,
respectively.

We now take the neutrino to be ultrarelativistic in the laboratory
frame where only a $B$-field exists. One can easily
derive the neutrino spin evolution by a Lorentz transformation to the
rest frame, where both a magnetic and electric field appear, and then
back to the laboratory frame. The well-known answer is that the
longitudinal component $B_\parallel$
contributes with a strength reduced by a factor of order $m/E$ caused by the
Lorentz transformations and can be neglected. The transverse component
$B_\perp$ leads to spin precession
with an effective magnetic moment $\mu_{\rm eff}=\mu+i\epsilon$.

In particular, this result means that the neutrino {\em electric\/} dipole
moment leads to a spin precession of ultrarelativistic neutrino in a
background {\em magnetic\/} field as first noted by Okun \cite{Okun:1986uf}.
Assuming the neutrino starts as a helicity eigenstate, the magnetic moment
leads to a precession in the plane perpendicular to ${\bm B}_\perp$. The
electric dipole moment, on the other hand, leads to a precession in the plane
spanned by ${\bm B}_\perp$ and the neutrino velocity ${\bm v}$. In other
words, the precession is around the direction ${\bm B}_\perp\times{\bm v}$
which is the electric field direction seen by the neutrino in its rest frame.
So for ultrarelativistic neutrinos, the real part of $\mu_{\rm eff}$ leads to a
``magnetic'' precession around ${\bm B}_\perp$, the imaginary part to an
``electric'' one around the direction of ${\bm B}_\perp\times{\bm v}$.

For Majorana neutrinos, the diagonal dipole moments vanish, whereas
the transition moments are~\cite{Shrock:1982sc}
\begin{subequations}\label{eq:transition-Majorana-moments}
\begin{eqnarray}
\mu_{ij}&=& (m_i+m_j)\frac{3e\sqrt{2}\GF}{(4\pi)^2}\,
i\,{\rm Im}\,F_{ij}\,,\\
i\epsilon_{ij}&=&(m_j-m_i)\frac{3e\sqrt{2}\GF}{(4\pi)^2}\,
{\rm Re}\,F_{ij}\,.
\end{eqnarray}
\end{subequations}
Notice that $F_{ij}$ is, strictly speaking, different from the Dirac
case because the leptonic mixing matrix for Majorana neutrinos
involves two additional Majorana phases beyond the Dirac phase.  In
general, $F_{ij}$ has both a real and an imaginary part. Therefore,
the spin precession will involve both the magnetic and electric type in
the sense explained earlier.

Overall, for both Dirac and Majorana neutrinos, the spin-flavor
transition $\nu_{i+}\leftrightarrow\nu_{j-}$ is proportional to $m_i$
whereas $\nu_{i-}\leftrightarrow\nu_{j+}$ is proportional to $m_j$,
i.e., they are different in contrast to what is usually written in the
literature on spin-flavor oscillations. The reason is that spin-flavor
transitions of ultarelativistic neutrinos are governed by an effective
transition moment of the form $\mu_{ij}\pm i\epsilon_{ij}$, and the
electric transition moment is frequently ignored in the literature.

In summary, the usual description in terms of dipole moments leads to
the same spin-flavor oscillation effects as found in our earlier
discussion in terms of a general background.

\subsection{Strong \boldmath{$B$}-field}

The discussion so far was limited to the weak-field case where only
the linear 
response to the applied $B$-field was considered and the effect can be lumped 
into a set of intrinsic neutrino magnetic and electric dipole and transition moments. 
However, if the $B$-field becomes very large, one needs to go beyond linear 
approximation in the dressed propagators of charged particles in Fig.~\ref{fig:blob1}.

Calculations of the neutrino self-energy under the influence of an external 
electromagnetic field beyond linear approximation
have a long history  \cite{Mckeon:1981ym,Borisov:1985ha,
Erdas:1990gy,Kuznetsov:2005tq,Kuznetsov:2007zza,BravoGarcia:2007spp,
Bhattacharya:2008px,Erdas:2009zh,Kuznetsov:2010sn,Dobrynina:2014rza,
Eminov:2016zqk}.
The probably most comprehensive study is
Ref.~\cite{Dobrynina:2014rza}. The linear $B$-regime is appropriate
for $eB<m_W^2$, i.e., below the $W$-critical field strength
of $B_W=m_W^2/e=1.09\times10^{24}$~G. Effects quadratic in $B$
include a modification of the dipole moments and the appearance of
an energy shift for massless neutrinos.

\subsection{Gauge invariance}

\begin{figure}[t!]
\includegraphics[width=0.80\columnwidth]{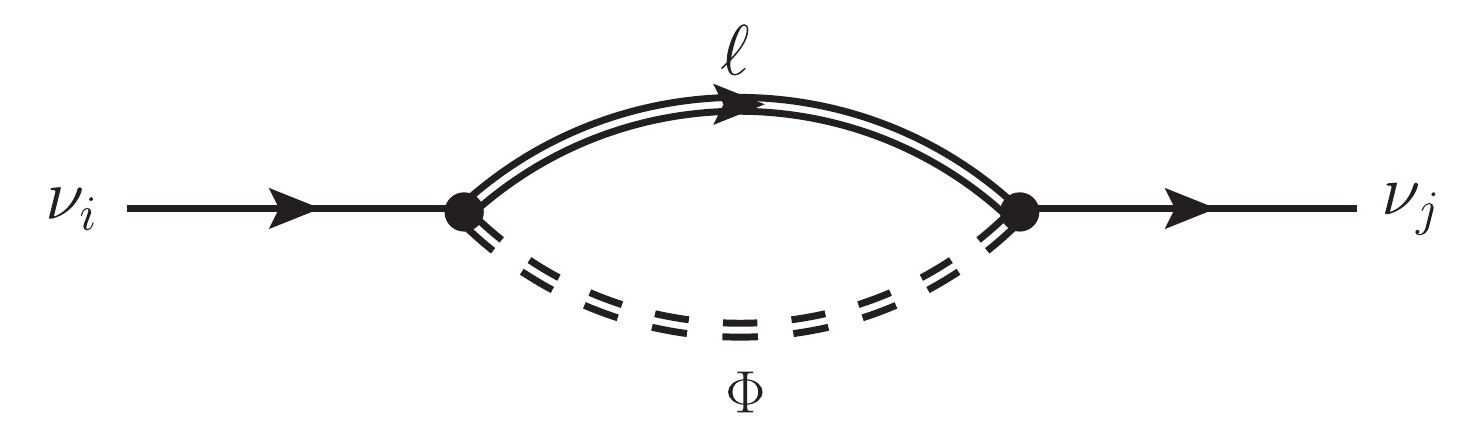}
\vskip2pt
\caption{\label{fig:diagram-with-Higgs}
Contribution of the charged Higgs (in unitary gauge it becomes
the longitudinal degree of freedom of the $W$-boson) to the neutrino
self-energy.}
\end{figure}

We have taken the attitude that the origin of neutrino masses and
neutrino interactions are unrelated and that the mechanism which provides
a Dirac or Majorana mass acts only on the external legs of the
self-energy graph in Fig.~\ref{fig:blob1}. In the classic calculation of
neutrino dipole moments by Shrock~\cite{Shrock:1982sc}, the author uses
unitary gauge and argues that physical results must be gauge invariant
so that we can use any gauge that is convenient. The
other classic calculation by Pal and Wolfenstein \cite{Pal:1981rm} was
performed in {'t}~Hooft--Feynman gauge, where the unphysical charged scalar
couples to the neutrino and charged lepton, see Fig.~\ref{fig:diagram-with-Higgs}.

For Dirac neutrinos and assuming their masses arise from ordinary
Yukawa couplings, gauge invariance of the magnetic dipole moment was
explicitly demonstrated in Ref.~\cite{Dobrynina:2014rza} using the
general $R_\xi$ gauge. In this gauge, the neutrino mass appears
on the level of an interaction vertex~\cite{Dobrynina:2014rza},
\begin{equation}
\label{eq:ChargedHiggsVertex}
\mathcal{L}= -\frac{g}{\sqrt{2}\,M_W}\left[(\bar{\nu}_i U^\dagger_{i\ell}K_{i\ell}\ell)\Phi^*+
\textrm{h.c.}\right]\,,
\end{equation}
where $K_{i\ell}=m_{\ell} R-m_i^{\rm D} L$ with $m_\ell$ the Dirac
mass of the charged lepton $\ell$.  Moreover, $m_i^{\rm D}$ is the
Dirac mass term for neutrino $i$, which coincides with the physical
neutrino mass, $m_i^{\rm D}=m_i$, if the Higgs interaction is the only
contribution.  To linear order in $m_i$ the resulting
contribution to the self-energy is proportional to $m_{i} L +
m_{j} R$. Sandwiching it between $\bar u_{i,s}$ and $u_{j,h}$
we find again that, to leading order in neutrino masses, the
transition amplitude $\nu_{i,-}\rightarrow \nu_{j,+}$ is proportional
to $m_j$, and the transition amplitude $\nu_{i,+}\rightarrow
\nu_{j,-}$ is proportional to $m_i$. Thus, the
neutrino Dirac mass is associated with the external neutrino line of
the ``wrong'' helicity. In other words, the  flavor structure of the 
transition amplitude produced by the diagram in Fig.~\ref{fig:diagram-with-Higgs} 
matches the one following from Fig.~\ref{fig:blob1}, which is a prerequisite 
for the gauge invariance of the total transition amplitude.

The propagators of the gauge and charged Higgs bosons depend on the
gauge parameter $\xi$, where the limit $\xi \rightarrow \infty$
corresponds to unitary gauge, whereas $\xi=1$ corresponds to
{'t}~Hooft--Feynman gauge. The leading contribution to the
self-energy, proportional to $\GF$, is independent of the gauge
choice. The next-to-leading contribution, proportional to $\GF/M^2_W$,
is given by a sum of terms, each of which is gauge dependent. However,
the gauge-dependent terms cancel out in the sum and the resulting
contribution to the self-energy is gauge independent as well
\cite{Dobrynina:2014rza}.

To perform an explicit calculation for Majorana neutrinos in an $R_\xi$ gauge one 
needs a concrete model. Pal and Wolfenstein \cite{Pal:1981rm} specifically
used a type-II see-saw model involving a Higgs triplet. In the see-saw type-I 
scenario, there are no additional Higgs fields, but  the right-handed neutrino 
acquires a large Majorana mass term $M_{\rm M}$. Diagonalizing the full mass 
matrix one finds that light states acquire masses of the order of 
$m \sim m_{\rm D}^2/M_{\rm M}$, while the heavy states get
masses of the order of $M_{\rm M}$. For the light mass eigenstates the Lagrangian has 
a form identical to the one in Eq.~\eqref{eq:ChargedHiggsVertex}, except that the Dirac mass
term $m^D_i$ in $K_{i\ell}$ is replaced by the physical neutrino mass $m_i$ and that the $3\times 3$
matrix $U$ is no longer unitary, but is a submatrix of the $6\times 6$ unitary mixing matrix. This 
implies, in particular, that the proof of the gauge invariance of the transition amplitude in the 
type-I see-saw scenario reduces to that for Dirac neutrinos.

\section{\label{sec:media}Polarized media}

Our initial interest was triggered by the observation that a medium
polarized by the $B$-field leads to helicity oscillations in addition
to those caused directly by the $B$-field. One crucial insight is that
particles and antiparticles with spins pointing in the same direction
have the same, not opposite, refractive effect. However, they have
opposite opposite magnetic moments.  Therefore, an ambient $B$-field
polarizes them in opposite directions and the particle and
antiparticle contributions to the induced spin polarization subtract.

From Lorentz covariance considerations one expects the $B$-field
induced axial-current contribution to the four-vector $b^\mu$ in the
self-energy Eq.~(\ref{eq:self-energy-form}) to have the structure
\smash{$u_\alpha \widetilde F^{\alpha\mu}$}, where $u$ is the four
velocity of the medium \cite{D'Olivo:1989cr}. In the remainder of this
section we assume to have a pure $B$-field in the rest frame of the
medium, so the structure is $b\propto (0,{\bm B})$. This expression is
to be compared with the $B$-field term in vacuum given in
Eq.~(\ref{eq:bfieldstructure}) where the role of $u$ is played by the
neutrino four momentum $k$.  While both terms are linear in ${\bm B}$,
their different structure explains their opposite sign change when the
test particle is switched to an antineutrino, as well as their very
different energy dependence.

We now discuss two examples which illustrate these
general points, i.e., an electron gas and a parti\-cle-antiparticle
symmetric medium.

\subsection{Electron gas}

The neutron and proton magnetic dipole moments are much smaller than those of
electrons, so typically the polarization effect is dominated by the electrons and
positrons \cite{Nunokawa:1997dp} on which we focus.
The electron spin-polarization contribution to neutrino
refraction is given by Eq.~(\ref{eq:matter-current}).
If the axial current represents the spin-polarization induced
by ${\bm B}$ it has only a spatial part ${\bm J}_\A\propto {\bm B}$ and the
four-vector contributing to the self-energy in Eq.~(\ref{eq:self-energy-form})
has the form $b=(0,{\bm b})$ with
\begin{equation}
{\bm b}=C_\A \sqrt{2}\GF\,{\bm J}_\A\,.
\end{equation}
In a bath of electrons and for $\nu_e$ as test particles,
$C_\A=1/2$,  whereas for $\nu_\mu$ and $\nu_\tau$ it is $-1/2$. For antineutrinos
as test particles, the signs reverse.

Probably the most complete derivation of the induced ${\bm J}_\A$ was
provided in Ref.~\cite{Nunokawa:1997dp} which we have independently
verified.  In an external $B$-field, the electrons and positrons
reside in Landau levels. Ignoring radiative corrections to the
electron magnetic moment, adjacent Landau levels with opposite spin
are degenerate and thus cancel in the expression for the
polarization. Overall only the lowest Landau level contributes which
corresponds to electrons or positrons moving along the $B$-field
direction. One therefore finds
\begin{equation}\label{eq:polarization-in-Bfield}
{\bm J}_\A=\frac{e{\bm B}}{2\pi^2}\int\limits_{-\infty}^{+\infty} \!\!dp
\left(\frac{1}{e^{(E-\mu_e)/T}+1}-\frac{1}{e^{(E+\mu_e)/T}+1}\right)\,,
\end{equation}
where $e\approx0.3028\ldots$ is the unit electric charge, $p$ the electron
or positron momentum along ${\bm B}$, $T$ the temperature,
$\mu_e$ the electron chemical potential,
and \smash{$E=\sqrt{m_e^2+p^2}$} the energy
of the lowest Landau level which is independent of $B$. The first term in
Eq.~(\ref{eq:polarization-in-Bfield}) is for electrons, the
second with reversed sign and reversed chemical potential for positrons.

In the limit of vanishing chemical potential (matter-anti\-ma\-t\-ter symmetric
plasma), ${\bm J}_\A$ vanishes as anticipated earlier. In the opposite limit
of a highly degenerate electron gas (vanishing temperature) one finds
\begin{equation}
{\bm J}_\A=\frac{e{\bm B}}{\pi^2}\sqrt{\mu_e^2-m_e^2}\,.
\end{equation}
These results apply to any field strength because only the lowest Landau
level contributes where the electrons move parallel to the $B$-field, but
the $B$-field influences the relationship between $\mu_e$ and
electron density. In particular, in the weak-field limit, we have
\smash{$\pF=\sqrt{\mu_e^2-m_e^2}$} with $\pF$ the electron Fermi momentum
corresponding to the electron density $n_e=\pF^3/3\pi^2$.
In this case, the induced energy shift for $\nu_e$  is
given by \cite{Semikoz:1987py, Nunokawa:1997dp}
\begin{equation}
{\sf H}^{\nu\nu}_{ii,--}=-\frac{e\sqrt{2}\GF\pF}{2\pi^2}\,B_\parallel\,,
\end{equation}
For  $\bar\nu_e$ the overall sign of the refractive terms reverses. For
the other flavors, the sign is opposite to the $\nu_e$ case.

For Dirac neutrinos, the spin-flip energy, i.e., the off-dia\-go\-nal
element in Eq.~(\ref{eq:Dirac-flavor}) is given by
\begin{equation}
{\sf H}^{\nu\nu}_{ii,+-}=-\frac{e\sqrt{2}\GF\pF}{2\pi^2}\,\frac{m_i}{2E}B_\perp
|U_{ei}|^2\,,
\end{equation}
to be compared with the expression following from the usual Dirac
dipole moment in Eq.~(\ref{eq:Dirac-dipole-moment}).
Therefore, the $B$-field induced spin-flip energy mediated by the degenerate
electron gas is $-|U_{ei}|^2 4\pF/3E_\nu$ times that
given by the vacuum dipole moment. In
an astrophysical setting, for example in a supernova core, typical neutrino
energies are not much smaller than a typical electron Fermi energy.
Therefore, the two contributions tend to be of similar magnitude.

For Majorana neutrinos, the electron-induced spin-flip energy is twice
that of the Dirac case. It is the only contribution because the
intrinsic Majorana dipole moment vanishes. Notice that for Majorana
neutrinos, the spin-flip energy is proportional to $m_i+\eta_b m_i$
and for the spin-polarization contribution $\eta_b=+1$, whereas for a
$B$-field in vacuum $\eta_b=-1$ as discussed earlier.  The general
transformation properties of in-medium neutrino electromagnetic
vertices were first studied in Ref.~\cite{Nieves:1989xg}. It was noted
that a Majorana neutrino can have a nonvanishing effective
electromagnetic vertex in a medium if the latter is not symmetric
between particles and antiparticles.

\subsection{Symmetric medium}

A matter-antimatter symmetric medium does not cause neutrino dispersion
in the framework of the low-energy current-current interaction.
In particular, there is no $B$-field induced spin polarization
as explained earlier
\cite{Semikoz:1987py, Semikoz:1994uy, Elmfors:1996gy, Nunokawa:1997dp}.
Moreover, there can not be an
effective Majorana dipole moment, whereas for the Dirac case
there is no objection from the general transformation
properties~\cite{Nieves:1989xg}. In other words, in such a medium one
has similar restrictions as in vacuum which can be seen as a CPT
symmetric ``medium'' of virtual particles.

However, in a CPT symmetric medium of real particles,
approximately realized in the early universe,
neutrinos suffer non-vanishing refraction~\cite{Notzold:1987ik}.
In the low-energy limit, it arises from an expansion of the gauge-boson
propagator, i.e., one needs to go beyond the local four-fermion interaction
model. The contribution of electrons and positrons was given
in Eq.~(\ref{eq:symmetric-bubble}).
The modified dispersion relation is the same for neutrinos and
antineutrinos.

In addition one may include a $B$-field which affects the electrons
and positrons. In the limit $m_e\ll T\ll m_W$ and $B\ll T^2$ one
finds for the $e^-e^+$ contribution to $\nu_e$ dispersion
\cite{Elmfors:1996gy, Erdas:1998uu, Kuznetsov:2005tq}\footnote{Notice
  that in Ref.~\cite{Erdas:1998uu} the metric $(-,+,+,+)$ is used.}
\begin{equation}
b^\mu=-\frac{e\sqrt{2}\GF T^2}{6 m_W^2}\,({\bm k}\cdot{\bm B},-k_0{\bm B})\,.
\end{equation}
It has a very similar structure to the effect of a $B$-field in vacuum
of Eq.~(\ref{eq:Bfield-vacuum}), except for a sign change of the spatial
part. In vacuum, the time-like and
longitudinal space-like parts nearly cancel so that $b_0-b_\parallel={\cal O}(m^2/E^2)$,
i.e., in vacuum a $B$-field does not cause dispersion even for active
neutrinos. In the present case, on the other hand, $b_0$ and $b_\parallel$ add up
so that for both $\nu_e$ and $\bar\nu_e$
\begin{equation}
{\sf H}^{\nu\nu}_{ii,--}=\frac{e\sqrt{2}\GF T^2}{3 m_W^2}\,{\bm k}\cdot{\bm B}\,.
\end{equation}
In addition, helicity conversion by neutrino masses occurs
in the now-familiar way, observing that $\eta_b=+1$.

\section{Conclusions}
\label{sec:conclusions}

It was recently recognized that a material medium which is not
isotropic in the laboratory frame instigates neutrino spin and
spin-flavor transitions caused by nonvanishing neutrino masses. It was
known for a long time that analogous effects arise from a $B$-field
due to neutrino dipole and transition moments, which in the simplest
case result from neutrino masses and their flavor mixing.

We have discussed these effects from the common perspective of
neutrino dispersion in a mean-field background, which could be a
material medium, a $B$-field, or both.  If neutrinos, be they Dirac or
Majorana particles, have only l.h.\ interactions (except for their
masses), the one-loop self-energy has a very simple form where for
ultrarelativistic neutrinos all dispersion effects are encoded in a
single four-vector which we called $b$.  It is the spatial part ${\bm
  b}$ which causes spin or spin-flavor conversion if it has a
nonvanishing component ${\bm b}_\perp$ transverse to the neutrino
direction of motion. The dependence on neutrino masses and mixing
parameters for all cases follows from the same structure.

Dirac neutrinos suffer helicity conversion, which for them is
simultaneously active-sterile conversion, for all types of anisotropic
backgrounds. On the other hand, Majorana neutrinos suffer helicity
conversion, which for them is simultaneously neutrino-antineutrino
conversion, only if $b$ changes sign when we switch from a test $\nu$
to a $\bar\nu$, which is the case for an ordinary background
medium. It is not true in a $B$-field, in a matter-antimatter
symmetric medium, or both. It is noteworthy that, while the helicity
of a single Majorana neutrino is not flipped by a $B$-field directly,
it is flipped by a normal medium polarized by the $B$-field.
Spin-flavor transitions arise for all types of anisotropic backgrounds
for both Dirac and Majorana neutrinos. The relative behavior of Dirac
and Majorana neutrinos is analogous to the well-known structure of
their dipole and transition moments.

Neutrino masses are extremely small. We do not know of a realistic
astrophysical or cosmological setting where the helicity evolution of
ordinary neutrinos would lead to observable consequences unless
the dipole or transition moments exceed those implied
by the masses. We still
think that our perspective on these issues adds some conceptual
clarity to the recent literature on neutrino spin evolution in
anisotropic media.

\section*{Acknowledgments}

We acknowledge partial support by the \emph{Deutsche
For\-schungs\-gemein\-schaft (DFG)\/} under Grant No.\ EXC-153 (Excellence
Cluster ``Universe''), and by the Research Executive Agency (REA) of the
European Union under Grant No.\ PITN-GA-2011-289442 (FP7 Initial Training
Network \emph{Invisibles}). A.D.\ acknowledges financial support by the \emph{Russian Foundation for
Basic Research} (Project No.\ 16-32-00066 {\cyr mol-a}), and by the
\emph{Dynasty Foundation}.

\end{document}